\documentclass[aps,prl,twocolumn,showpacs,floatfix,nofootinbib,preprintnumbers,superscriptaddress,amsmath,amssymb]{revtex4-1}

\usepackage{graphicx}% Include figure files
\usepackage{dcolumn}% Align table columns on decimal point
\usepackage{bm}% bold math
\usepackage{hyperref}% add hypertext capabilities
\usepackage[mathlines]{lineno}% Enable numbering of text and display math
\usepackage{times}
\usepackage{datetime}
\newcommand{\hfbax}{\sc hfb-ax}

\newcommand{\rr} {\boldsymbol{r}}

\begin{document}

\title{ Probing Surface Quantum Flows in Deformed Pygmy Dipole Modes }% Force line breaks with \\

\author{Kai Wang}
\affiliation{State Key Laboratory of Nuclear
Physics and Technology, School of Physics, Peking University,  Beijing 100871, China}

\author{M. Kortelainen}
\affiliation{Department of Physics, P.O. Box 35 (YFL), University of Jyvaskyla, FI-40014 Jyvaskyla, Finland}
\affiliation{Helsinki Institute of Physics, P.O. Box 64, FI-00014 University of Helsinki, Finland}

\author{J.C. Pei}
\email{peij@pku.edu.cn}
\affiliation{State Key Laboratory of Nuclear
Physics and Technology, School of Physics, Peking University,  Beijing 100871, China}

\begin{abstract}

In order to explore the nature of collective modes in weakly bound nuclei, we have investigated deformation effects
and surface flow patterns of isovector dipole modes in a shape-coexisting nucleus $^{40}$Mg. The calculations were done in
a fully self-consistent continuum finite-amplitude Quasiparticle Random Phase Approximation (QRPA)
in a large deformed spatial mesh.
An unexpected result of pygmy and giant dipole modes having disproportionate deformation splittings in strength functions was obtained.
Furthermore, the transition current densities demonstrate that the long-sought core-halo oscillation in pygmy resonances is collective and compressional,
corresponding to the lowest excitation energy and the simplest quantum flow topology.
Our calculations show that surface flow patterns become more complicated as excitation energies increase.

\end{abstract}


%\today
\maketitle

%\tableofcontents

\emph{Introduction.}---
The atomic nuclei are in the evolution from
few-body to many-body quantum systems and could exhibit amazing collective phenomena.
In particular, when nuclei have large unbalanced neutron to proton ratios (i.e., a large isospin asymmetry) and associated significant charge-neutral
surfaces, new coherent excitation modes are expected to appear~\cite{pyg-theo,pyg-expt}.
The so-called ``soft" or ``pygmy" dipole resonance (PDR), corresponding to the relative oscillation between
the core and the skin or halo from the hydrodynamical point of view~\cite{pygmy},  are especially intriguing.
They are relevant to rich physics aspects such as the neutron skin,  equation of state,
as well as enhanced neutron capture rates in astrophysical \textit{r}-process~\cite{pyg-theo}.
Extensive experimental measurements have been performed to study the PDR, although mainly in less exotic nuclei~\cite{pyg-expt,ca48}.
Theoretically, however, it is of ultimate interests to explore the nature of
pygmy resonances in weakly bound nuclei close to drip lines, in contrast to giant resonances.

It is expected that both pygmy and giant resonances can have non-degenerate modes induced by deformation effects.
Since the deformed halo structures have been proposed~\cite{misu,sgzhou,pei2013}, it is desirable to identify decoupled halo-core shapes, through
comparative studies of anisotropic $K$-splittings in pygmy and giant dipole resonances (GDR). However, detailed splitting behaviors of PDR have not been studied yet.
Furthermore, in PDR of weakly-bound nuclei, whether there exists a collective core-halo  oscillation,
is still a long-standing question, which can be perceived directly via transition currents.
The associated irrotational superfluid flow could be an intuitive PDR mode or a compressional mode or a toridal mode, according to Refs.~\cite{vretenar,Nesterenko,Papakonstantinou}.
It is essential to
study the flow patterns with broken spatial symmetries even for spherical nuclei,
because by imposing symmetries the internal motions and flow patterns can be notably different~\cite{ring}.
The study of nuclear surface flow features should be of broad interests, considering the interesting flows in cold atomic gases~\cite{bulgac2} and quark matter~\cite{mayg}.

The suitable microscopic tool for aforementioned studies is the fully self-consistent deformed continuum
quasiparticle random-phase-approximation (QRPA), which takes into account weak-binding effects.
The fully self-consistency is important for elimination of spurious states, in order to obtain meaningful low-lying states.
In addition, low-lying excitations can be sensitive to pairing and continuum effects~\cite{matsuo,khan}.
Due to numerically challenging fully self-consistent treatment of deformed continuum QRPA, the earlier studies have
been usually limited to spherical symmetry~\cite{matsuo,khan}.
Recently, the finite-amplitude-method QRPA (FAM-QRPA) has been proposed~\cite{fam}, which
allows to solve the QRPA problem iteratively, being more efficient compared to the conventional matrix-QRPA approach.
In the present work, we demonstrate that the multipole excitations in deformed weakly bound nuclei can
be studied with unprecedented numerical accuracy by adopting the FAM-QRPA approach in a large spatial mesh.

In this Rapid Communication, we aim to study deformation splittings  and visualize the surface
flow patterns in PDR of the weakly bound deformed $^{40}$Mg. The calculations are done in
the fully self-consistent, coordinate space, deformed continuum FAM-QRPA framework.
$^{40}$Mg is the last experimentally observed magnesium isotope~\cite{nature-mg} with a $N$=28 magic neutron number,
but with a well-established prolate-oblate shape coexistence~\cite{Terasaki,egido}.
We stress that such a shape-coexistence scenario is ideal to comparatively analyze deformation effects in pygmy resonances.
Compared to earlier studies of $^{40}$Mg~\cite{yoshida}, our calculations employ a large spatial mesh
and, consequently, can provide details of PDR which have never been revealed before.

\begin{figure}[t]
  \includegraphics[width=0.46\textwidth]{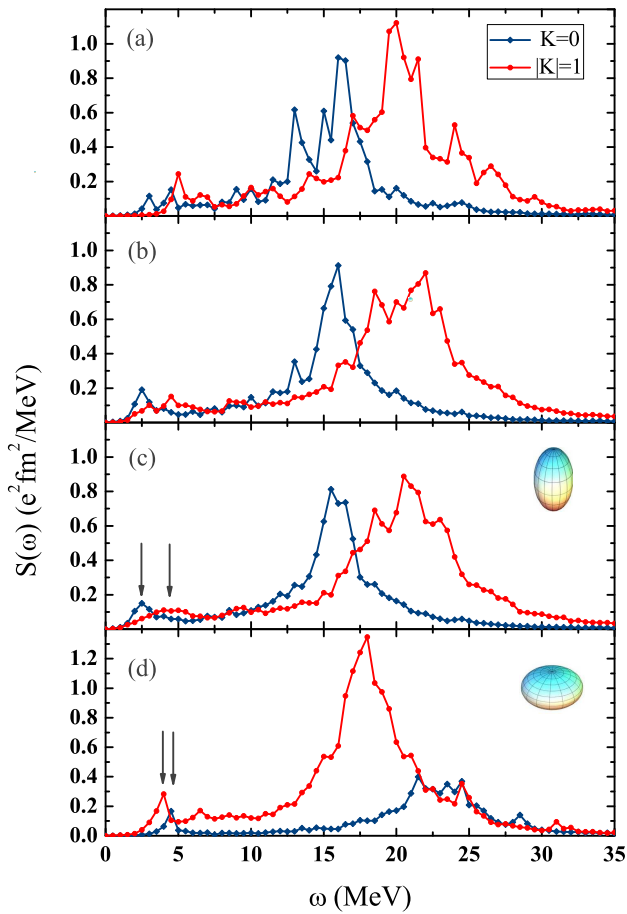}\\
  \caption{(Color online) Transition strength functions of isovector dipole resonances in the shape-coexisting $^{40}$Mg as a function of excitation energy,
  calculated with different box sizes.
    (a) the prolate shape with a box size of 12\,fm;    (b) the prolate shape with a box size of 21\,fm.
    (c) the prolate shape with a box size of 27.6\,fm;  (d) the oblate shape with a box size of 27.6\,fm.  }
  \label{fig1}
\end{figure}

\emph{Method.}---
Here we describe the development of the FAM-QRPA approach for multipole collective excitations based on Hartree-Fock-Bogoliubov (HFB) solutions in a large,
axially symmetric coordinate-space mesh. In deformed weakly-bound nuclei, there exists a subtle interplay
among the surface deformation, surface diffuseness and continuum coupling.
Therefore, it is crucial to prepare very precise ground-state HFB solutions as a starting point for reliable descriptions of collective excitations.
The HFB equation is solved with the computer code {\hfbax}~\cite{Pei08} within a large two-dimensional coordinate-space mesh,
based on B-spline techniques for axially symmetric deformed nuclei~\cite{teran}.
For calculations employing large box sizes and small lattice spacings,
the discretized continuum spectra would be very dense, providing good resolutions of quasiparticle resonances and continuum~\cite{pei2013}.
Note that the exact, fully self-consistent, treatment of continuum in deformed cases~\cite{matsuo2009} is rare
and has not been used for QRPA calculations yet.

For the particle-hole interaction channel, a recently adjusted extended SLy4 force for light nuclei is adopted~\cite{xiong},
including an additional density dependent term.
For the particle-particle channel, a density dependent pairing interaction, $V_{0}[1-\eta(\rho(\rr)/\rho_0(\rr))^{\gamma}]$,
is used~\cite{pastore}.  With a pairing window of 60\,MeV, the pairing force parameters are taken as $V_0=-448.3\,{\rm MeV\,fm^3}$,
$\eta$=0.8 and $\gamma$=0.7, so that pairing gaps in both stable and very neutron-rich nuclei can be properly described.
The resulted pairing gaps are between those from mixed and surface types of pairing.
A pure surface pairing interaction may overestimate pairing correlations in nuclei far from stability~\cite{chongqi}.
Detailed results with SLy4 and extended-SLy4 forces, and different pairing interactions
are shown in the Supplemental Material~\cite{supple}.

The next step is to combine the FAM-QRPA method with the deformed coordinate-space HFB approach.
Recently, there has been a lot of new developments with the FAM-QRPA method~\cite{fam-o}.
In our previous work, we implemented FAM-QRPA in axially deformed coordinate-spaces for monopole transitions~\cite{pei14}.
Only very recently, the FAM-QRPA scheme with an arbitrary multipole operator has been realized~\cite{markus15}.

In our FAM-QRPA calculations, to maintain full self-consistency, the full quasiparticle basis and all time-odd terms are included.
Note that the widely used canonical basis approach to apply truncations can undermine descriptions of halo tails~\cite{dobaczewski96}.
To study the fine structures of pygmy resonances in present work, an imaginary part of the excitation frequency $\omega$, for smoothing resonances,
is taken to be 0.25\,MeV, which is smaller than the usually adopted 0.5\,MeV.
The maximum angular momentum $z$-projection limit for quasiparticle states is taken as $\Omega_{\rm max}=39/2$.
The storage of all wave functions takes about 17 Gigabytes when a box size of 27.6\,fm is adopted.
We solve the non-linear multipole FAM-QRPA equation iteratively with the modified Broyden method.
For each excitation frequency $\omega$ point,
the calculation employs the OpenMP shared memory parallel scheme. For different frequencies,
the MPI distributed parallel scheme is adopted. Consequently, the FAM-QRPA approach can be efficiently
implemented by using this hybrid parallel scheme.

\begin{figure}[t]
  \includegraphics[width=0.48\textwidth]{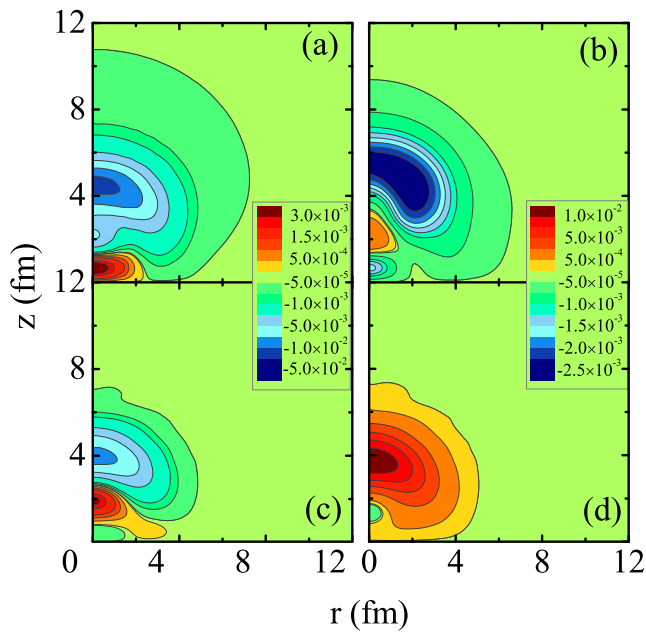}\\
  \caption{(Color online) The $K$=0 transition density distributions of pygmy and giant dipole resonances in $^{40}$Mg with the prolate shape, are displayed
  in the cylinder coordinate space as  $\delta\rho$(\textit{r}, \textit{z}).
  (a) for neutrons of PDR, (b) for neutrons of GDR, (c) for protons of PDR, (d) for protons of GDR.   }
  \label{fig2}
\end{figure}

\emph{Results.}---We have performed calculations for the coexisting prolate and oblate shapes in $^{40}$Mg,
which has a very soft potential energy surface~\cite{Terasaki}.
The prolate shape ($\beta_2=0.39$)
is a weakly bound ground state with a neutron Fermi energy of $\lambda_{\rm n}=-0.33$\,MeV, which is more reasonable
compared to the standard SLy4 result of $\lambda_{\rm n}=-0.52$\,MeV, considering that $^{42}$Mg and $^{39}$Na have not been experimentally observed yet.
The oblate minimum ($\beta_2=-0.32$) is energetically about 1.9\,MeV above the prolate minimum, but has a lower Fermi energy
of $\lambda_{\rm n}=-0.79$\,MeV. Note that the isotone $^{42}$Si has
a well-deformed oblate ground state~\cite{si42}.
This shape competition is also reflected in superfluid properties.
The prolate shape has a neutron pairing gap of $\Delta_{\rm n}=1.23$\,MeV (while $\Delta_{\rm p}=0$),
and the oblate shape has a proton pairing gap of $\Delta_{\rm p}=0.98$\,MeV (while $\Delta_{\rm n}=0$).

The electric isovector dipole resonances at the prolate and oblate local energy minima can be self-consistently obtained without spurious states,
since, within FAM-QRPA, the isovector dipole operator does not excite spurious center-of-mass mode.
The isovector PDR is a natural probe of relative surface oscillations and can be also related to the nuclear photo absorption cross section.
The isoscalar PDR, which suffers from spurious states~\cite{terasaki05,arteaga1}, is not investigated presently.
In Fig.~\ref{fig1}, the calculated FAM-QRPA transition strength functions of $K=0$ and $\vert K\vert=1$ (with sum of $K=1$ and $K=-1$) cases are shown.
Our implementation has been benchmarked with the calculation of $^{24}$Mg (see details in Ref.~\cite{supple}).
It is known that soft monopole resonances are mainly due to continuum effects,
rather than due to single-particle structures~\cite{pei14}, and this also holds for soft multipole resonances.
To see the role of the accurate treatment of continuum and halo extensions,
the transition strengths have been calculated with box sizes of 12\,fm, 21\,fm and 27.6\,fm, respectively.
We can see that,  within a small box, the continuum discretization is insufficient
and several false peaks appear.  Even with a box size of 21\,fm, a false peak at 13\,MeV is present.
Moreover, it can be seen that pygmy resonances are fragmented and less coherent with smaller box sizes.

With calculations employing a large box size of 27.6\,fm, the obtained transition strengths clearly demonstrate smoothed pygmy resonances
and deformation-induced splittings, as shown in Fig.~\ref{fig1}~(c,d).
It is known that the anisotropic splitting in the dipole transition strength is approximately proportional to
the centroid excitation energy and the deformation, based on a hydrodynamic model~\cite{bertsch}.
The microscopic RPA calculations have confirmed that the GDR splitting depends linearly on  the deformation~\cite{arteaga2}.
The results show that prolate and oblate shapes have similar GDR splittings ($\delta_E \sim 5$\,MeV).
Considering different centroid energies, the estimated PDR deformation splittings should be around 0.95\,MeV for the
prolate shape and 1.05\,MeV for the oblate shape.
However, we see the PDR deformation splitting of the prolate shape ($\delta_E=1.4$\,MeV) is significantly larger compared to the
expected value whereas the oblate case ($\delta_E=0.45$\,MeV) is on the contrary.
Calculated density distributions show no core-halo shape decoupling in $^{40}$Mg (see ~\cite{supple}).
Obviously, the mechanism for the PDR deformation splitting differs from the GDR splitting.
We have tested that obtained deformation splittings are not sensitive to pairing gaps.
Therefore, we speculate that the pygmy deformation splitting is related to not only the static shape but also significant dynamical surface effects.
It will be very helpful to look for the PDR deformation splitting by high-resolution experimental measurements of deformed neutron rich nuclei.
The dominating $\vert K\vert=1$ mode in the oblate case and also the total cross section, differ notably from the prolate case.

\begin{figure}[t]
  \includegraphics[width=0.48\textwidth]{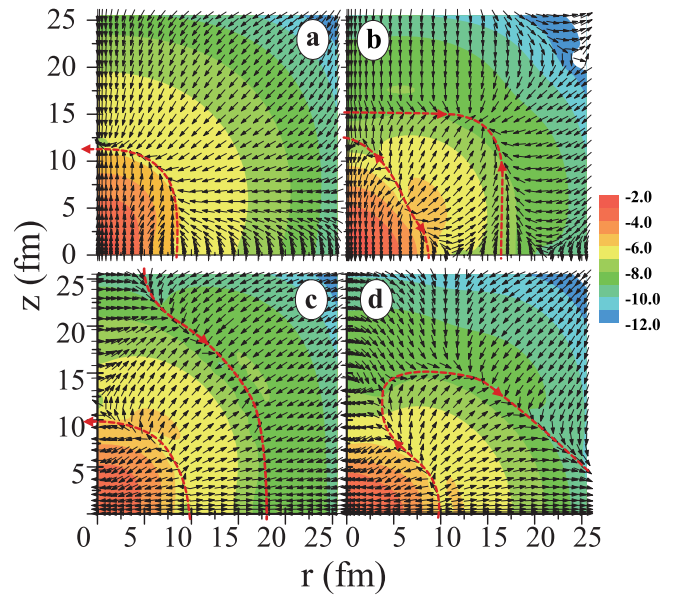}\\
  \caption{(Color online) The neutron transition current density in the cylinder coordinate space, $\delta \vec{j}(r, z,\phi=0)$, of the PDR in $^{40}$Mg.
  The color scales denote the logarithm of the current strength and the arrows denote the flow direction.
  (a) $K$=0 mode of the prolate shape, (b) $K$=0 mode of the oblate shape, (c) $K$=1 mode of the prolate shape, (d) $K$=1 mode of the oblate shape.
  The flow boundary lines are given for guiding eyes.     }
  \label{fig3}
\end{figure}

To understand the different collective nature between pygmy and giant resonances, we have further studied the
transition density $\delta \rho$  and the transition current density $\delta \vec{j}$ at resonance peaks. Note that the smoothing of resonances
has been included in solving the FAM-QRPA equation.
Fig.~\ref{fig2} displays the imaginary part of transition density distributions of $K=0$ modes of the prolate shape.
We see that in the pygmy resonance the neutron-proton transition densities have basically an in-phase pattern
whereas the giant resonance has a characteristic pattern of opposite phases.
This is consistent with the relative core-halo motion in pygmy resonances.
This in-phase pattern has been pointed out in the literature and is related to strong isoscalar PDR, e.g., in Refs.~\cite{yoshida,paar,paar2}.
However, a contrary interpretation was obtained with RPA calculations in Ref.~\cite{prc87.014324}.
For the weakly bound $^{40}$Mg, we see the neutron transition density has a much larger spatial extension than that of protons.
The excessive neutron surface is the main source of the PDR transition strength.

Fig.~\ref{fig3} displays the imaginary part of the neutron transition current density $\delta\vec{ j}$ on a cylinder $(r,z)$-plane with
azimuthal angle $\phi =0$.
The transition current density can be used as a direct probe of the collective nature of pygmy resonances.
The very small current fields at surfaces actually correspond to considerable large velocity fields~\cite{supple}.
To compare with the superfluid prolate case, the pairing strength in calculations of oblate current flows has been increased by $5\%$.
This does not affect to the resulting PDR deformation splitting.

In Fig.~\ref{fig3} (a), the current flow of the prolate shape $K=0$ PDR illustrates a very clear compressional pygmy mode,
which is exactly the long-sought collective surface-core oscillation~\cite{Nesterenko}.
There is a flow-in pole at $z=12$\,fm along the $z$-axis, connected with the corresponding flow-out pole at $z=-12$\,fm due to the
reflection symmetry, showing a typical dipole structure.
The boundary between centripetal inward and outward flows appears at a distance of 12\,fm from the center.
Such a large boundary distance is related to a very soft excitation (2.5\,MeV).
The $K=0$ PDR of the prolate shape has the simplest flow topology, associated with the lowest energy (2.5\,MeV).
On the other hand, the flows corresponding to Fig.~\ref{fig3}~(b, c, d) all have two boundaries with a rebound wave and
have similar excitation energies ($4\thicksim4.5$\,MeV).
This energy dependence of flow patterns may be related to the disproportionate $K$-splittings in PDR as shown in Fig.\ref{fig1}.
We see the compressional flow patterns are characterized by boundary lines, in analog to the topological winding numbers, but without any internal vorticity.
Calculations with different box sizes up to 34 fm have been performed~\cite{supple}, which confirm that
the compressional PDR pattern is a collective quantum phenomenon, independent of box sizes.
The dominance of surface compressional flows in weakly-bound nuclei is not a surprise since the dilute nuclear surface has
a very small incompressibility and, consequently, a remarkable soft monopole mode~\cite{pei14}.

It is worth to mention the flow pattern of the $K$=1 PDR of the oblate shape in Fig.~\ref{fig3}~(d).
We see that two boundaries are connected and a flow circulation is generated.
There have been several studies~\cite{vretenar,Papakonstantinou,Nesterenko} pointing out the possible existence of a
toroidal mode in neutron-rich nuclei, which may be favorable in the isoscalar PDR at higher energies.
In our results, with the standard isovector dipole operator, the flow circulation is a synthetic pattern from sideward compressional flows (not an explicit toroidal mode),
demonstrating the complexity of flow patterns due to the finite-size deformation effect and the possible interplay with the toroidal mode.
By contrast, this circulation is not seen in the centripetal $K$=1 PDR mode of the prolate shape.

\begin{figure}[t]
  \includegraphics[width=0.48\textwidth]{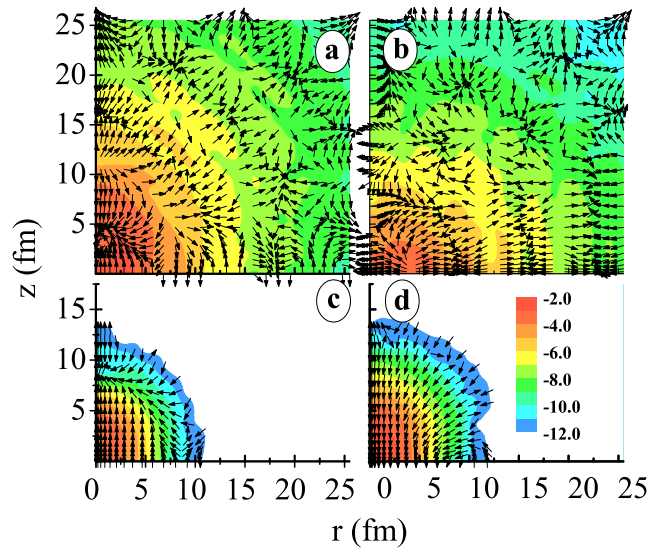}\\
  \caption{(Color online) Transition currents similar to Fig.~\ref{fig3} but for
  (a) the neutron $K$=0 prolate GDR, (b) the neutron $K$=1 oblate GDR,
  (c) the proton $K$=0 prolate PDR, (d) the proton $K$=0 prolate GDR. }
  \label{fig4}
\end{figure}

In Fig.~\ref{fig4}~(a, b), the neutron surface current flows of the $K$=0 prolate and $K$=1 oblate GDR
are much more complex compared to the PDR case.
Nevertheless, the flow pattern of the $K=0$ prolate GDR has several compressional ring structures.
The compressional structures in GDR have been demonstrated again by calculations with different box sizes~\cite{supple}.
In addition, there is an orderly displacement wave at box boundaries.
There are some flow poles and disorders as a consequence of flow interferences, depending on box sizes.
As the excitation energy increases, distortions in the nuclear inner part appear in the flow.
Generally, it is difficult to identify geometrical shape effects in flow patterns in giant resonances.

Fig.~\ref{fig4}~(c,d) show the proton flows of the prolate pygmy and giant resonances.
The proton flows have much smaller spatial extensions and show no distortions in either case.
The in-phase flow pattern of PDR is very evident as shown in Fig.~\ref{fig3}~(a) and Fig.~\ref{fig4}~(c).
A possible way to investigate the charge-neutral currents at outer nuclear surfaces, connected to novel scissors
and twist modes, is through magnetic transitions~\cite{heyde}.

\emph{Summary.}--- With the newly developed fully self-consistent continuum deformed FAM-QRPA
approach in a large spatial mesh, we have investigated the distinct collective nature of pygmy and giant dipole
resonances of $^{40}$Mg associated with weak-binding effects.
Firstly, the deformation splitting in pygmy resonances was found to be disproportional to the GDR splitting.
This disproportion is not due to the static core-halo shape decoupling effect, implying considerable dynamical surface effects at low energies.
Furthermore, the transition current flows illustrate very different topologies associated with excitation energies and
geometrical shapes in pygmy resonances.
The long-sought surface-core PDR oscillation is collective and compressional, corresponding to the simplest quantum flow topology, which occurs at a large
distance boundary as the lowest $K$=0 mode of the ground-state prolate shape.
The flow pattern has been demonstrated to be a robust quantum  phenomenon by calculations with different box sizes.
The surface flow patterns were seen to become more complicated as excitation energies increase.
The analysis of these transition current flows can provide an insight
of the pygmy resonance, in addition to the analysis of transition densities.
The present study demonstrates that anisotropic large coordinate-space calculations
are essential for exploring soft excitations in weakly bound nuclei.

\begin{acknowledgments}
Useful comments by W. Nazarewicz and F.R. Xu are gratefully acknowledged.
This work was supported by the National Natural Science Foundation of China under Grants No.11522538, 11375016,11235001.
This work was also supported (M.K.) by the Academy of Finland under the Centre of Excellence
Programme 2012-2017 (Nuclear and Accelerator Based Physics Programme at JYFL) and FIDIPRO program.
We also acknowledge that computations in this work were performed in Tianhe-1A
located at Tianjin and Tianhe-2
located at Guangzhou.
\end{acknowledgments}

\nocite{*}

%\bibliography{yinu}% Produces the bibliography via BibTeX.

\end{document}